\documentclass[fleqn]{annalen}
 \usepackage{epsf}
\begin{document}
\newcommand{\volume}{8}              
\newcommand{\xyear}{1999}            
\newcommand{\issue}{7--9}               
\newcommand{\recdate}{29 July 1999}  
\newcommand{\revdate}{dd.mm.yyyy}    
\newcommand{\revnum}{0}              
\newcommand{\accdate}{dd.mm.yyyy}    
\newcommand{\coeditor}{ue}           
\newcommand{\firstpage}{593}           
\newcommand{\lastpage}{602}           
\setcounter{page}{\firstpage}        
\newcommand{\keywords}{quantum phase transitions, quantum magnets, critical behavior}
\newcommand{\PACS}{75.20.En; 75.45.+j; 64.60.Kw}
\newcommand{\shorttitle}{T. Vojta et al., Quantum critical behavior of
                         itinerant ferromagnets}
\title{Quantum critical behavior of itinerant ferromagnets}
\author{Thomas Vojta$^{1,2}$, D. Belitz$^{2}$,
    T. R. Kirkpatrick$^{3}$, and R. Narayanan$^{2}$}
\newcommand{\address}
{$^1$Institut f{\"u}r Physik, Technische Universit\"at Chemnitz, D-09107
Chemnitz, FRG\\
$^2$Dept. of Physics and Materials Science Institute,
                                      University of Oregon, Eugene, OR 97403\\
$^3$Institute for Physical Science and Technology, and Department of Physics,\\
     University of Maryland, College Park, MD 20742\\}
\newcommand{\email}{\tt vojta@physik.tu-chemnitz.de}
\maketitle
\def\tr{{\rm tr}\,}
\def\Tr{{\rm Tr}\,}
\def\sgn{{\rm sgn\,}}
\def\b{\bibitem}
\def\boldphi{\mbox{\boldmath $\phi$}}
\def\boldvarphi{\mbox{\boldmath $\varphi$}}
\begin{abstract}
We investigate the quantum phase transition of itinerant ferromagnets. It is
shown that correlation effects in the underlying itinerant electron system
lead to singularities in the order parameter field theory that result in an
effective long-range interaction between the spin fluctuations. This
interaction turns out to be generically {\em antiferromagnetic} for clean
systems. In disordered systems analogous correlation effects lead to even
stronger singularities. The resulting long-range interaction is, however,
generically ferromagnetic.

We discuss two possibilities for the ferromagnetic quantum phase transition.
In clean systems, the transition is generically of first order, as is
experimentally observed in MnSi. However, under certain conditions the
transition may be continuous with non-mean field critical behavior.
In disordered systems, one finds
a very rich phase diagram showing first order and continuous phase
transitions and several multicritical points.

\end{abstract}

\section{Introduction}
\label{sec:I}

Quantum phase transitions are phase transitions that occur at zero temperature
as a function of some non-thermal
control
parameter. The fluctuations that drive these
transition are of quantum nature rather than thermal in origin. Among the
transitions that have been investigated are various metal-insulator
transitions, the superconductor-insulator transition in thin metal films, and
a variety of magnetic phase transitions. Quantum phase transitions have
attracted considerable attention in recent years, in particular since they are
believed to be at the heart of some of the most exciting discoveries in modern
condensed matter physics, such as the localization problem, various magnetic
phenomena, the quantum Hall effects, and high-temperature
superconductivity \cite{qhehtc}.

One of the most obvious examples of a quantum phase transition
is the transition from a paramagnetic to a ferromagnetic metal
that occurs
as a function of the exchange coupling between the electron spins.
The experimentally best studied example of such a transition
is probably
provided by the pressure-tuned transition in MnSi \cite{Lonzarich}.
MnSi belongs to the class of so-called
nearly or weakly ferromagnetic materials. This group of metals,
consisting of transition metals and their compounds such as ZrZn$_2$,
TiBe$_2$, Ni$_3$Al, and YCo$_2$ in addition to MnSi are
characterized by strongly enhanced spin fluctuations. Thus, their
ground state is close to a ferromagnetic instability which makes them good
candidates for actually reaching the ferromagnetic quantum phase transition
experimentally
by changing the chemical composition or applying pressure.

\begin{figure}[t]
  \epsfxsize=7.5cm
  \centerline{\epsffile{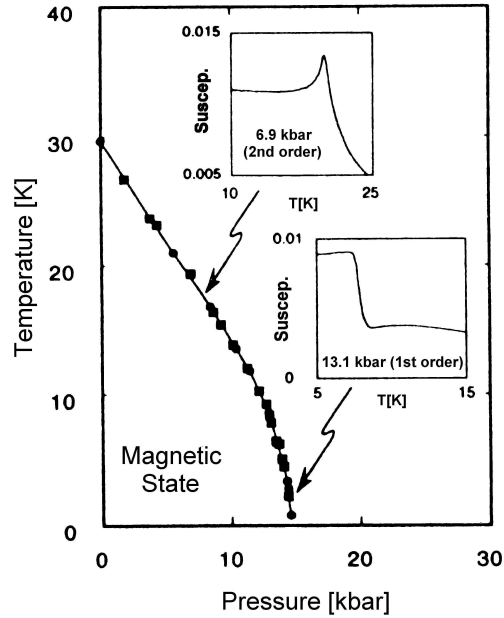}}
  \caption{Phase diagram of MnSi. The insets show the behavior of
   the susceptibility close to the transition.
  (After \protect\cite{Lonzarich}).}
  \label{fig:MnSi}
\end{figure}

At ambient pressure MnSi is paramagnetic for temperatures larger
than $T_c=30$\,K. Below $T_c$ it orders magnetically. The order is,
however, not exactly ferromagnetic but a long-wavelength
(190\,\AA)  helical spin spiral along the (111) direction of the
crystal. The ordering wavelength depends only weakly on
the temperature, but a homogeneous magnetic field of about 0.6\,T suppresses
the spiral and leads to ferromagnetic order.
One of the most remarkable findings about the magnetic phase transition
in MnSi is that it changes from continuous to first order with decreasing
temperature as is shown in Fig.\ \ref{fig:MnSi}.
Specifically, in an experiment carried out at low
pressure (corresponding to a comparatively high transition
temperature) the susceptibility shows a pronounced maximum
at the transition, reminiscent of the singularity expected
from a continuous phase transition.
In contrast, in an experiment at a pressure very close to (but still smaller
than) the critical pressure the susceptibility does not show any sign of a
divergence at the phase transition. Instead, it displays a finite
discontinuity
suggestive of a first-order phase transition.

A related set of experiments is devoted to a phenomenon called
{\em itinerant electron metamagnetism}. Here a high magnetic field is
applied to a nearly ferromagnetic material such as
Co(Se$_{1-x}$S$_x$)$_2$ or Y(Co$_{1-x}$Al$_x$)$_2$ \cite{metamag}.
At a certain field strength the magnetization
of the sample shows a pronounced jump. This can easily be explained
if we assume that the free energy
as a function of the magnetization has the triple-well structure
characteristic of the vicinity of a first-order phase transition.
In zero field the side minima must have a larger free energy than the
center minimum (since the material is paramagnetic in zero field).
The magnetic
field essentially just ''tilts'' the free energy function. If one of the side
minima becomes lower than the center (paramagnetic) one, the magnetization
jumps.

In the literature the first-order transition in MnSi at low temperatures
as well
as the itinerant electron
meta\-magnetism have been attributed to sharp structures in the electronic
density of states close to the Fermi energy which stem from the
band structure of the particular material. These
structures in the density of states can lead to a negative
quartic coefficient in a magnetic Landau theory and thus to the
above mentioned triple-well structure.

In this paper it will be shown, however, that the two phenomena are generic
since they are rooted in the universal many-body physics underlying the
transition. Therefore, they
are predicted to
occur for all nearly or weakly ferromagnetic
materials irrespective of special structures in the density of states. In this
paper we will emphasize the basic physics behind these findings, a technical
derivation can be found elsewhere
\cite{us_clean,us_dirty,us_chi_s,us_firstord}.

The paper is organized as follows: In Sec.\ \ref{sec:II} we sketch
the derivation of
an order parameter field theory for the ferromagnetic quantum phase
transition of itinerant electrons, starting from a microscopic
description of an interacting electron system and analyze its
properties.
In Sec.\ \ref{sec:III} we compare the possible scenarios for the
quantum phase transition in detail, while Sec.\ \ref{sec:IV} is
devoted to the influence of quenched disorder. We conclude
in Sec.\ \ref{sec:V}.

\section{Order parameter field theory}
\label{sec:II}

In a pioneering paper
that
was the first application of
the modern theory of critical phenomena to a quantum phase transition,
Hertz \cite{Hertz} derived an order parameter field theory for
the ferromagnetic quantum phase transition by considering a simple
model of itinerant electrons
that interact only via the exchange interaction in the particle-hole
spin-triplet channel. Hertz analyzed this order parameter field theory
by means of renormalization group (RG) methods. He found a continuous
phase transition whose critical behavior in the
physical dimensions $d=3$ and $d=2$ is
mean field-like,
since the dynamical critical exponent $z=3$ decreases the upper critical
dimension from $d_c^+ =4$ for the classical case to $d_c^+=1$ in the
quantum case. Despite the somewhat artificial
nature of this model, it was believed for a long time that the qualitative
features of Hertz's analysis, in particular the fact that there is
mean field-like
critical behavior for all $d>1$, apply to real itinerant
quantum ferromagnets as well.

Here we will show, however, that this belief is mistaken.
The properties of the
ferromagnetic quantum phase transition are
much more complicated since the magnetization
couples to additional, non-critical soft modes in the electronic system.
Mathematically, this renders the conventional Landau-Ginzburg-Wilson approach
invalid since an expansion of the free energy in powers of the order
parameter does not exist. Physically, the additional soft modes
lead to an effective
long-range interaction between the order parameter fluctuations.
This long-range interaction, in turn, can change the character
of the transition from a
continuous transition with mean-field exponents to either
a continuous transition with non-trivial
(non-mean field)
critical behavior or even to a first
order transition like in MnSi.

The derivation of our theory \cite{us_clean} follows Hertz \cite{Hertz}
in
spirit, but the technical details are considerably different. We consider a
$d$-dimensional continuum model of interacting electrons, and pay particular
attention to the particle-hole spin-triplet contribution \cite{AGD} to the
interaction term in the action, $S^t_{\rm int}$, whose (repulsive) coupling
constant we denote by $\Gamma_t$. Writing only the latter explicitly, and
denoting the spin density by ${\bf n}_s$, the action reads,
\begin{equation}
S = S_0 + S_{\rm int}^t
  = S_0 + (\Gamma_t/2) \int dx\ {\bf n}_s(x)\cdot{\bf n}_s(x)\quad,
\label{eq:6}
\end{equation}
where $S_0$ contains all contributions
to the action other than $S_{\rm int}^t$. In particular, it contains the
particle-hole spin-singlet and particle-particle interactions,
which will be important for what
follows.\footnote{We note, however, that even in Hertz's original model,
  where $S_0$ describes free electrons, interaction terms get generated
  upon renormalization. The traditional mean-field results therefore
  are not correct for this model either.}
$\int dx = \int d{\bf x} \int_0^{1/T} d\tau$, and we use
a 4-vector notation $x = ({\bf x}, \tau)$, with ${\bf x}$ a vector
in real space, and $\tau$ imaginary time. Following Hertz, we perform a
Hubbard-Stratonovich decoupling of $S_{\rm int}^t$ by introducing a
classical vector field ${\bf M}(x)$ with components $M^i$ that couples to
${\bf n}_s(x)$ and whose
average is proportional to the magnetization, and we
integrate out all fermionic degrees of freedom.
We obtain the partition function $Z$ in the form
\begin{equation}
Z = e^{-F_0/T} \int D[{\bf M}]\,\exp\bigl[-\Phi[{\bf M}]\bigr]\quad,
\label{eq:7a}
\end{equation}
where $F_0$ is the non-critical part of the free energy.
The Landau-Ginzburg-Wilson (LGW) functional $\Phi$ reads
\begin{eqnarray}
\Phi[{\bf M}]&=& {1\over 2} \int dx\,dy\ {1\over \Gamma_t}\ \delta(x-y)\
{\bf M}(x)\cdot {\bf M}(y) \label{eq:7b} \\
&&+ \sum_{n=2}^{\infty}a_n \int dx_1\,\ldots\,dx_n\ \chi^{(n)}_{i_1\ldots i_n}
  (x_1,\ldots,x_n)
M^{i_1}(x_1)\,\ldots\,M^{i_n}(x_n)\ ,
\nonumber
\end{eqnarray}
where $a_n = (-1)^{n+1}/n!$. The coefficients
$\chi^{(n)}$ in (\ref{eq:7b}) are connected n-point spin density
correlation functions of a
reference system with action
$S_0$ \cite{us_dirty, Hertz}.
The particle-hole spin-triplet interaction $\Gamma_t$ is missing in the bare
reference system, but a nonzero $\Gamma_t$ is generated perturbatively by the
particle-particle interaction contained in $S_0$.
The reference system then has all of the characteristics of
the full action $S$,
except that it must not undergo a phase transition lest the separation of
modes that is implicit in our singling out $S_{\rm int}^t$ for the
decoupling procedure breaks down.

$\chi^{(2)}$ is the spin susceptibility of the
reference system. Performing a Fourier
transform from $x=({\bf x},\tau)$ to $q=({\bf q},\Omega)$ with wave vector
${\bf q}$ and Matsubara frequency $\Omega$, we have for small ${\bf q}$ and
$\Omega$,
\begin{equation}
\chi^{(2)}({\bf q},\Omega)
           = \chi_0({\bf q})[1 - \vert\Omega\vert/\vert{\bf q}\vert]\quad,
\label{eq:8a}
\end{equation}
where ${\bf q}$ and $\Omega$ are being measured in suitable units, and
$\chi_0({\bf q})$ is the static spin susceptibility of the reference system.
If we take the susceptibility to be that of a non-interacting Fermi gas,
$\chi_0({\bf q}\rightarrow 0) = c_0 - c_2 {\bf q}^2$, we obtain
Hertz's
theory. However, in a real Fermi liquid,
the static spin susceptibility at $T=0$ is a
non-analytic function of ${\bf q}$ since the magnetization modes couple to
additional soft modes, viz. particle-hole excitations in the spin-triplet
channel with a ballistic dispersion relation \cite{us_chi_s}. This
non-analyticity is crucial for the non-trivial physics of the itinerant
ferromagnets discussed in this paper. For small wave vectors the static
susceptibility is of the form
\begin{equation}
\chi_0({\bf q}\rightarrow 0) = c_0 - c_{d-1}\vert{\bf q}\vert^{d-1}
                                                - c_2 {\bf q}^2 \quad.
\label{eq:8b}
\end{equation}
Here $c_0$, $c_{d-1}$ and $c_2$ are constants. This holds for $1<d<3$; in
$d=3$ the non-analyticity is of the form $-\tilde c_2{\bf q}^2 \ln\vert{1/ \bf
q}\vert$. Note that all these singularities only exist at zero temperature and
in zero magnetic field since both a finite temperature or a magnetic field
gives the particle-hole excitations a mass.

Using (\ref{eq:8a},\ref{eq:8b}),
and with $\int_q = \sum_{\bf q} T\sum_{i\Omega}$,
the Gaussian part of $\Phi$ can be written,
\begin{equation}
\Phi^{(2)}[{\bf M}] = \int_q {\bf M}(q)\bigl[t_0
                       + c_{d-1}\vert{\bf q}\vert^{d-1} + c_2{\bf q}^2
                + c_\Omega \vert\Omega\vert/\vert{\bf q}\vert\bigr]\,
                                                 {\bf M}(-q)\quad.
\label{eq:9}
\end{equation}
Here $t_0 = 1 - \Gamma_t\chi^{(2)}({\bf q}\rightarrow 0,\omega_n = 0)$
is the bare distance from the critical point, and $c_\Omega$
is another constant.

For the same physical reasons for which the non-analyticity occurs in
(\ref{eq:8b}), the higher coefficients $\chi^{(n)}$ ($n>2$) in (\ref{eq:7b})
are in general not finite at zero frequencies and wave numbers. Generally, the
coefficient of $\vert{\bf M}\vert^{n}$ in $\Phi$ for $\vert{\bf
p}\vert\rightarrow 0$ behaves like $\chi^{(n)} = v^{(n)} \vert{\bf
p}\vert^{d+1-n}$. This implies that $\Phi$ contains a non-analyticity which in
our expansion takes the form of a power series in $\vert{\bf
M}\vert^2/\vert{\bf p}\vert^2$. Consequently, the free energy functional
(\ref{eq:7b})
is mathematically ill defined. However, we will nonetheless be
able to extract a considerable amount of information.

The sign of the non-analyticity in the Gaussian term merits some attention
since it will be responsible for the qualitative features of the ferromagnetic
quantum phase transition. Perturbation theory to second order in $\Gamma_t$
yields $c_{d-1}<0$ \cite{us_clean,us_chi_s}. This is the generic case, and it
is consistent with the
well-known
notion that correlation effects in general
decrease the effective Stoner coupling \cite{White}. However, Ref.\
\cite{us_chi_s} has given some possible mechanisms for $c_{d-1}$ to be
positive at least in some materials.

\section{Phase transition scenarios}
\label{sec:III}

Depending on the sign of the non-analyticity in the
Gaussian term (\ref{eq:9}) of the free energy functional the properties
of the ferromagnetic quantum phase transition will be qualitatively
different.

We first discuss the generic case of $c_{d-1}<0$. Here the free energy reduces
with increasing $q$ from zero which implies that a continuous transition to a
ferromagnetic state is impossible at zero temperature. Two possible scenarios
for the phase transition arise for $c_{d-1}<0$. The first scenario is based on
the observation that a finite thermodynamic magnetization $m=\langle |{\bf
M}(x)|\rangle$, which acts similarly to a magnetic field, cuts
off
the singularities in the coefficients of the order parameter field theory.
Therefore, the non-analyticity in $\chi^{(2)}$ leads to an analogous
non-analyticity in the magnetic equation of state, which takes the form
\begin{eqnarray}
 t m - v_d m^d + u m^3 &=& H \qquad (d<3)~, \\
 t m - v_3 m^3 \ln(1/m) + u m^3 &=& H \qquad (d=3)~,
\label{eq:equofstate}
\end{eqnarray}
where $t$ tunes the transition and $u, v_d$ and $v_3$ are positive constants.
$H$ denotes the external magnetic field.
This equation of state describes a first-order phase
transition since the next-to-leading term for small $m$ has
a
negative sign.
We have investigated
this scenario in some detail \cite{us_firstord}.
Since the non-analyticities
in $\chi^{(2)}$ and the equation of state are cut off by
a finite temperature, the transition will be of first order
at very low $T$ but turn second order at higher temperatures.
The two regimes are separated by a tricritical point.

The second possible scenario for the quantum phase transition
arising if
$c_{d-1}<0$ is that the ground state of the system will not be ferromagnetic
but instead a spin-density wave at finite {\bf q}. This scenario has not been
studied in much detail so far, but work is in progress. It is tempting to
interpret the spiral ordering in MnSi as a signature of this finite-$q$
instability. This is, however, not very likely since a finite-$q$ instability
caused by our long-range interaction will be strongly temperature dependent
due to the temperature cutoff of the singularities. As mentioned above,
experimentally the ordering wave vector is essentially temperature
independent. Further work will be necessary to decide which of the two
possible scenarios, viz. a first-order ferromagnetic transition or a
continuous transition to modulated magnetic order, is realized under what
conditions. Moreover, let us point out, that in $d=3$ the non-analyticity is
only a logarithmic correction and would hence manifest itself only as a phase
transition at exponentially small temperatures, and exponentially large length
scales. Thus, it may well be unobservable experimentally for some materials.

We now turn to the second case, $c_{d-1}>0$ which can happen, if one of the
conditions discussed in Ref.\ \cite{us_chi_s} is fulfilled. In this case the
self-generated long-range interaction is ferromagnetic. Consequently, the
ferromagnetic quantum phase transition will be a conventional second order
phase transition, which can be analyzed by standard renormalization group
methods. A tree level analysis shows that the Gaussian theory is sufficient
for dimensions $d>d_c^+=1$ since all higher order terms are irrelevant. We are
therefore able to obtain the critical behavior exactly, yet due to the
long-range interaction it is not
mean field-like.
The results of this analysis \cite{us_clean} can be summarized as
follows. The equation of state close to the critical point reads
\begin{eqnarray}
 t m + v_d m^d + u m^3 &=& H \qquad (d<3)~, \label{eq:1a}\\
 t m + v_3 m^3 \ln(1/m) + u m^3 &=& H \qquad (d=3)~,
\label{eq:1b}
\end{eqnarray}
Again, $u$ and $v$ are positive constants. Note the different sign
of the non-analytic term compared to (\ref{eq:equofstate}).
From (\ref{eq:1a},\ref{eq:1b})
one obtains the critical exponents $\beta$ and $\delta$, defined by
$m \sim t^{\beta}$ and $m \sim H^{1/\delta}$, respectively, at $T=0$.
For $\beta$ and $\delta$, and for
the correlation length exponent
$\nu$, the order parameter susceptibility exponent $\eta$, and the
dynamical exponent $z$, we find
\begin{equation}
\beta = \nu = 1/(d-1),\quad \eta = 3-d,\quad \delta = z = d\quad,\quad
(1<d<3)\quad,
\label{eq:2}
\end{equation}
and $\beta=\nu=1/2$, $\eta=0$, $\delta=z=3$ for $d>3$. These exponents
`lock into' mean-field values at $d=3$, but have nontrivial values for
$d<3$. In $d=3$, there are logarithmic
corrections to power-law scaling. Eqs.\ (\ref{eq:1a},\ref{eq:1b}) apply
to $T=0$.
At finite temperature, we find homogeneity laws for $m$, and for the
magnetic susceptibility, $\chi_m$,
\begin{equation}
m(t,T,H) = b^{-\beta/\nu} m(tb^{1/\nu}, Tb^{\phi/\nu}, Hb^{\delta\beta/\nu})
                                                                    \quad,
\label{eq:3a}
\end{equation}
\begin{equation}
\chi_m(t,T,H) = b^{\gamma/\nu} \chi_m(tb^{1/\nu}, Tb^{\phi/\nu},
                           Hb^{\delta\beta/\nu})\quad,
\label{eq:3b}
\end{equation}
where $b$ is an arbitrary scale factor. The exponent $\gamma$, defined by
$\chi_m \sim t^{-\gamma}$ at $T=H=0$ and the crossover exponent $\phi$
that describes the crossover
from the quantum to the classical Heisenberg fixed point (FP) are given by
\begin{equation}
\gamma = \beta (\delta - 1) = 1 \quad,\quad \phi = \nu \quad,
\label{eq:4}
\end{equation}
for all $d>1$. Notice that the temperature dependence of the
magnetization is {\em not}
given by the dynamical exponent. However, $z$ controls the temperature
dependence of the specific heat coefficient, $\gamma_V = c_V/T$, which
has a scale dimension of zero for all $d$, and logarithmic corrections
to scaling for all $d<3$ \cite{Wegner},
\begin{equation}
\gamma_V(t,T,H) = \Theta(3-d)\,\ln b +
           \gamma_V(tb^{1/\nu}, Tb^z, Hb^{\delta\beta/\nu})\quad .
\label{eq:5}
\end{equation}
Eqs. (\ref{eq:1a}) -- (\ref{eq:5}) represent the exact critical
behavior of itinerant quantum Heisenberg ferromagnets for all
$d>1$ with the exception of $d=3$, where additional logarithmic
corrections to scaling appear \cite{us_clean}.

\section{Influence of disorder}
\label{sec:IV}

\begin{figure}[t]
  \epsfxsize=\textwidth
  \centerline{\epsffile{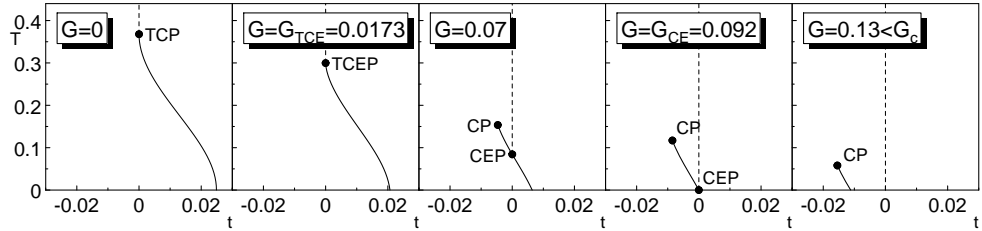}}
  \caption{Phase diagrams
  of disordered itinerant ferromagnets in the $T$-$t$-plane showing first
  order (solid) and second order (dashed) transitions. $G$ is a dimensionless
  measure of the disorder, and CP, CEP, TCP, and TCEP denote critical points,
  critical endpoints, tricritical points and tricritical endpoints,
  respectively. See Ref.
  {\protect\cite{us_firstord}} for more details.}
  \label{fig:first}
\end{figure}

In this section we briefly
discuss the influence of
quenched non-magnetic disorder on the ferromagnetic quantum
phase transition. An approach along the lines of the one
for the clean case sketched in Sec. \ref{sec:II} has been
developed in Ref. \cite{us_dirty}, and
the resulting effective theory
is very similar. Again, the magnetization couples to additional
soft modes (here with diffusive dynamics) which leads to an
effective long-range interaction. The singularities are even
stronger than in the clean case, but they have the opposite sign
so that the long-range interaction is generically ferromagnetic.
Thus in the presence of disorder there will be a
competition between the ballistic and diffusive singularities.
and the temperature which cuts off both.
For weak disorder the first-order transition will survive,
while larger disorder leads to a continuous transition.
As shown in Ref. \cite{us_firstord},
the phase diagram becomes very rich, showing
several multicritical points and even
regions with metamagnetic behavior (see Fig. \ref{fig:first}).
The properties of the continuous quantum phase transition
occurring for stronger disorder can again be analyzed by
standard renormalization group methods.
It turns out that as in the clean case the Gaussian
theory is sufficient since all higher order terms
are irrelevant. The resulting critical exponents are
\begin{equation}
\gamma = 1\quad,
\label{gamma}
\end{equation}
for all $d>2$,
\begin{equation}
\nu = 1/(d-2)\quad,\quad\eta = 4-d\quad,\quad z = d\quad,
\label{nuetaz}
\end{equation}
for $2<d<4$, while $\nu=1/2$, $\eta=0$, and $z=4$ for $d>4$.
In addition to $d=4$, $d=6$ also plays the role of an upper
critical dimension, and one has
\begin{equation}
\beta = 2/(d-2)\quad,\quad\delta = d/2\quad,
\label{betadelta}
\end{equation}
for $2<d<6$, while $\beta=1/2$, $\delta=3$ for $d>6$.

An important problem  in disordered systems that has attracted
a lot of attention within the last years is the influence of
rare disorder fluctuations on the properties of the phase
transition. In the conventional perturbative approach
\cite{us_dirty} the rare regions are neglected.
We have developed a theory which includes the rare regions into
a renormalization group approach. This theory is discussed
in a separate paper in this volume \cite{hh_islands}.
Here we quote only the final result for the itinerant ferromagnet:
Due to the effective long-range interaction which stabilizes the
Gaussian theory, rare regions do {\em not} change the properties
of the ferromagnetic quantum phase transition (in contrast to,
e.g., itinerant antiferromagnets).

\section{Conclusions}
\label{sec:V}
To summarize, we have discussed the ferromagnetic quantum phase
transition in itinerant electron systems. It has been shown that
the critical magnetization modes couple to additional, non-critical
soft modes in the electronic systems which results in an effective
long-range interaction between the magnetization fluctuations.
We have discussed several possible scenarios for the ferromagnetic
quantum phase transition.
In clean systems the generic scenario is a first-order phase transition.
This  provides us with a complete explanation for the nature of
the transitions observed in MnSi,  which in Ref.\ \cite{Lonzarich}
were attributed to a band structure feature characteristic of MnSi.
While
this feature may well be sufficient to make the transition in MnSi of first
order, the present theory leads to the surprising prediction that the first
order transition is {\em generic}, and thus should be present in other weak
clean itinerant ferromagnets as well. Our theory further
predicts in detail how the
first order transition will be suppressed by quenched disorder. Observations
of such a suppression, or lack thereof, would be very interesting for
corroborating or refuting the theory. Semi-quantitatively, the theory
predicts that the $T$ region that shows a first order transition will
be largest for strongly correlated systems.
Conversely, since the dependence of the
tricritical temperature on the system parameters is exponential,
in some, or even many, systems the first order transition may take place
only at very low temperatures. This may explain why in ZrZn$_2$
no first order transition has been observed \cite{Lonzarich}, although the
experiment does not seem to rule out a weakly first order
transition \cite{Pfleiderer}.

Under certain conditions, the transition in clean itinerant ferromagnets
can also be of second order. These conditions have been formulated
mathematically. Their precise physical or experimental nature has
not been investigated so far, but it is known that strong
correlations are a necessary condition. If these conditions are fulfilled,
then the critical behavior is known exactly, albeit it is not mean field-like.
If the system contains sufficiently strong quenched disorder, the transition is always of second order,
and the non-mean field-like critical behavior has been determined exactly.

Finally, it has been shown that the presence of rare regions or local moments
in the disordered case does not change these results, in contrast to, e.g.,
the case of itinerant antiferromagnets. The reason lies in the effective
long-range interaction between the order parameter fluctuations in the
ferromagnetic case, which is sufficient to suppress all fluctuations at the
critical point, including the static disorder fluctuations responsible for
rare regions. The same suppression of fluctuations is also the reason behind
our ability to determine the critical behavior for bulk systems, and indeed
for any dimension, exactly.

\vspace*{0.25cm} \baselineskip=10pt{\small \noindent
We thank G. Lonzarich and C. Pfleiderer for helpful discussions, and the
Aspen Center for Physics for hospitality during the completion of this
paper. This work was supported by
the NSF under grant Nos. DMR--98--70597 and DMR--99--75259,
and by the DFG  (SFB 393/C2).}

\end{document}